\documentclass{svproc}
\usepackage{url}

\usepackage{graphicx}
\usepackage{array}
\newcolumntype{P}[1]{>{\raggedright\arraybackslash}p{#1}}

\begin{document}
\mainmatter              % start of a contribution
\title{Using network science to quantify economic disruptions in regional input-output networks}

\titlerunning{Disruptions on input-output networks}  % abbreviated title (for running head)
%                                     also used for the TOC unless
%                                     \toctitle is used
%
\author{Emily P. Harvey\inst{1,3} \and Dion R.J. O'Neale\inst{2,3}}
\authorrunning{Harvey and O'Neale} % abbreviated author list (for running head)
%
%%%% list of authors for the TOC (use if author list has to be modified)
\tocauthor{Emily P. Harvey, Dion R.J. O'Neale}
\institute{M.E Research, Takapuna, Auckland, NZ\\
\email{emily@me.co.nz},\\
\and
Physics Department, University of Auckland, Private Bag 92019, Auckland, NZ\\
\and
Te P\={u}naha Matatini, Centre of Research Excellence, NZ}

\maketitle              % typeset the title of the contribution

\begin{abstract}
%The abstract should summarise the contents of the paper
%using at least 70 and at most 150 words. It will be set in 9-point
%font size and be inset 1.0 cm from the right and left margins.
%There will be two blank lines before and after the Abstract. \dots

Input-output (IO) tables provide a standardised way of looking at interconnections between all industries in an economy, and are often used to estimate the impact of disruptions or shocks on economies.
IO tables can be thought of as networks -- with the nodes being different industries and the edges being the flows between them. 
We develop a network-based analysis to consider a multi-regional IO network at regional and sub-regional level within a country.
We calculate both traditional matrix-based IO measures (e.g. ‘multipliers’) and new network theory-based measures, and contrast these measures with the results of a disruption model applied to the same IO data. 
We find that path-based measures, such as betweenness centrality, give a good indication of flow-on economic disruption, while eigenvector-type centrality measures give results comparable to traditional IO multipliers. We also show the effects of treating IO networks at different levels of spatial aggregation.
%147 words

% We would like to encourage you to list your keywords within
% the abstract section using the \keywords{...} command.
\end{abstract}
\section{Introduction}
Economic disruptions such as those due to natural hazards have a large impact on local and global economies. There is evidence that the flow-on impacts of disruptions will have an increasing impact as the world becomes more globalised and inter-connected \cite{Acemoglu2012,ContrerasFagiolo2014}. 
In order to build resilience and prioritise investment to mitigate impacts, it is crucial to identify key industry sectors and regions that play a role in amplifying (or dampening) the flow-on impacts of disruptions or shocks. 
Internationally, studies of flow-on impacts of disruptions on economic systems have most commonly been based on the data from input-output (IO) tables, which are readily available, at least at a national, and often also at regional levels.  Many years of research has gone into using IO tables in economic impact analysis \cite{MillerBlair2009} and a common approach is to use ‘multipliers’, based on linear algebra matrix formulations, to estimate the indirect impacts of a change in demand (or supply) for an industry. In response to natural hazard events, the most popular approach has been inoperability input-output models, which are based on IO tables with a small tweak \cite{DiestzenbacherMiller2015}.

The standard linear algebra approaches of `IO analysis' have some fundamental limitations. Recently the IO tables have been thought of as networks, with the nodes being the different industries and the edges being the flows between the industries. This has enabled network science techniques to be used to attempt to identify crucial nodes (industry sectors) within an economy and other industry structures. Existing work in this field began with calculating the properties of IO networks \cite{Blochl2011,Cerina2015,McNerney2013}, and is now beginning to investigate the propagation of shocks on the networks
\cite{ContrerasFagiolo2014,Liang2016,McNerney2013,Xu2011}.

In this work we seek to consider the connections between industries as a network, to determine what information and insights the structure of the network can give us. We consider a multi-regional IO network at local (Territorial Authority) level within the Waikato Region in New Zealand, and  calculate both matrix-based IO measures (e.g. `multipliers') and network theory-based measures at this higher spatial resolution. We compare these network-based measure with results from a disruption model applied to the same IO data, which gives us further information about disruption impacts. 

Research to date has considered the world IO network, looking at flows within and between each country, or looking at a single country of interest, but regions are often heterogeneous and impacts can occur locally. By comparing and contrasting the analysis at both Regional and Local spatial resolutions, we are also able to investigate the impact of spatial resolution on the results obtained.

\section{Background and Data}
\subsection{Level of spatial and industry aggregation}
In this work, our starting data is a Multi-regional Input-Output table (MRIO) which partitions the 10 Territorial Authorities (TAs) in the Waikato Region into separate subregions and breaks the rest of New Zealand into ‘North of the Waikato Region’ (Auckland and Northland) and ‘South of the Waikato Region’ (all other Regions). This gives us 12 different spatial regions which span a large range of sizes, both geographically and economically. The number of industry sectors is aggregated to 106, which is the maximum allowed from the reference data used to construct the IO tables. 

\subsection{Economic network setup} 
IO networks are weighted, directed networks, where the weighting indicates the size (\$) of inter-industry flows, and the direction depends on which industry the flow is from and to. In this work we have 1272 industry nodes (12 regions and 106 sectors) in the network. In addition to the flows between industry sectors, the IO tables we use include inputs from four Value Added categories (labour and capital inputs, taxes/subsidies on products and production) and Imports, and outputs to three Final Demand categories (household and government consumption, capital formation) and Exports. This adds another three nodes to each of the 12 `regions' for Final Demand, four nodes for Value Added at whole country level, and a node each for Imports and Exports.

It is a requirement of IO tables that flows in and out must balance, so this places restrictions on the network. Additionally, because the nodes are the grouping of all industries of the same `sector' in the TA, self-links are possible.  There are some other features of IO networks that it is worth noting; one of which is that they are very dense (nearly-complete) with most industries having connections with most other industries. These features mean that a lot of the standard approximations and simplifications for weighted, directed network analysis are often not able to be applied \cite{McNerney2009}.

\section{Method}
\subsection{Network analysis}
The `importance' or `influence' of a node in a network is estimated in network science approaches using a range of different \emph{centrality} measures.  We can broadly characterise most of these measures as direct industry measures, e.g.~strength and diversity, neighbourhood (eigenvector-type) measures, e.g.~eigenvector centrality and PageRank, or overall network structure (path-based) measures, e.g.~betweenness and closeness. In this work we analyse the network using a range of different centrality measures, included those that have been identified as potentially important in economic network, see Table \ref{tab:centralitymeasures}. Where possible we consider the network as a weighted, directed network, with self-loops, but not all algorithms allow for this.

Having calculated this selection of centrality measures, we use Kendall's $\tau$ \cite{Kendall} to calculate the correlation between the importance rankings of the industries between any two centrality measures. 

{\renewcommand{\arraystretch}{1.2}%
\begin{table}[htbp]
\centering
\begin{tabular}{P{3cm}|@{\hspace{1em}} p{7.5cm} @{\hspace{1em}}|l}
\hline
Vertex centrality measure    & Description   & Package    \\
\hline
Strength  (in/out/total)      &  The total production (inputs/outputs) \$ of the industry.    & igraph \cite{igraph}    \\
Diversity   & An (undirected) measure of the rarity or commonness of the industry   & igraph \cite{igraph}    \\
Eigenvector centrality       &   How connected the industry is to other industries and/or high-strength industries.       & igraph \cite{igraph}    \\
PageRank algorithm      &     How connected the industry is to other industries and to high-strength industries.      & igraph \cite{igraph}    \\
Kleinberg's authority centrality  &  Eigenvector centrality variant - high authority score means that the industry is connected to many `hubs'      & igraph \cite{igraph}    \\
Kleinberg's hub centrality    &  Eigenvector centrality variant - high hub score means that the industry is connected to many `authorities'         & igraph \cite{igraph}    \\
SALSA hub centrality&  An (unweighted \& undirected) measure which is a normalised version of the Kleinberg hub score                  & centiserve  \cite{centiserve}\\
SALSA authority centrality  &  An (unweighted \& undirected) measure which is a normalised version of the Kleinberg authority score    & centiserve\cite{centiserve} \\
Bonacich power centrality     &  Eigenvector centrality variant - How connected the industry is to either well-connected or isolated industries ($\beta$=0.5,1,2,10)     & igraph \cite{igraph}    \\
Bonacich alpha centrality  & Eigenvector centrality variant - which considers endogenous and exogenous factors ($\alpha$= 1,2)      & igraph \cite{igraph}    \\
Betweenness     & The number of shortest paths that go through the industry, where the weighting affects the `distance' between all industries.     & igraph \cite{igraph}    \\
Burt's constraint    & Consider which industries act as `brokers' between gaps in the network structure  & igraph \cite{igraph}    \\
Closeness centrality (in/out/total) & Steps required to access every other industry downstream of a given industry (out) or to access the given industry from every other upstream industry (in). & igraph \cite{igraph}    \\
Latora closeness centrality (in/out)   & Variant of closeness centrality which allows for consideration of disconnected components    & centiserve \cite{centiserve}\\
Eccentricity  (in/out)    & The shortest path distance (weighted) to any other industry in the system (out) or from any other industry (in)    & igraph \cite{igraph}    \\
Diffusion degree (in/out)      &  Path-based measure approximating the influence to downstream (out) or from upstream (in) industries under diffusion along the weighted network & centiserve  \cite{centiserve}\\
ClusterRank centrality  &  An (unweighted) measure which takes into account the local clustering of industries as well as the number of downstream industries   & centiserve  \cite{centiserve}\\
S-core decomposition             & Weighted version of the k-core decomposition       & brainGraph \cite{brainGraph}        
\end{tabular}
\vspace*{5mm}
\caption{List of network centrality measures calculated}
\label{tab:centralitymeasures}
\end{table}

\subsection{Multiplier analysis} 

\subsubsection*{IO multipliers}
The main technique used for quantifying economic impacts of a disaster (for example a natural hazard such as an earthquake or volcanic eruption) is known as Inoperability Input-Output Model (IIM) or the Dynamic IIM as a time varying extension \cite{DiestzenbacherMiller2015}. In this model, the inoperability of industries is assumed to follow a smooth logistic curve from the disaster induced loss of productive capacity back to full capacity over a specified recovery period. The direct loss of production due to industry inoperability is calculated and used to modify the final demand by the same amount; that is,  if production halved from \$20,000 to \$10,000 then the final demand vector for that industry in that region would be reduced by \$10,000. Then the flow on impacts from this reduced demand would be calculated using the Leontief inverse \cite{MillerBlair2009}. This continues through time, until full operability is restored.

In this work we calculate both Type I (industry to industry spending only) multipliers and Type II (including household spending and labour income) multipliers, for the whole multi-regional IO table, following \cite{MillerBlair2009}.

\subsubsection*{Disruption multipliers} 
There are many issues with this IIM approach \cite{Oosterhaven2017}, including that it can lead to double counting and not only inaccurate quantitative results, but more importantly it can lead to inaccurate rankings for prioritisation of industries.  Harvey \emph{et al.} \cite{Harvey2019} have instead developed a dynamic  model that propagates short-term (days to weeks) disruptions through the multi-regional IO network. Using this model `disruption multipliers' can be calculated by disrupting one industry at a time in each region and working out the ratio of direct effects to flow-on (indirect) effects throughout the whole of New Zealand.

\subsection{Comparing Network Centralities with IO and Disruption Multipliers}
We use Kendall’s $\tau$ \cite{Kendall} to calculate the correlation between industry rankings based on the multiplier measures compared to the centrality measures.

\subsection{Comparing spatial aggregation} 
In parallel with this, we also construct a network at the level of the Waikato Region (not separated at TA level) and same the `North of the Waikato Region' and `South of the Waikato Region' (3 network regions) to investigate the impact of spatial aggregation of IO tables on the economic multipliers and the network properties.

\section{Results}

\subsection{Network centrality measures and their correlations}
\begin{figure}[h]
\caption{Centrality measure correlations for industries in the ten TAs in the Waikato Region, colour scale from blue ($\tau$=-0.84) to red ($\tau$=1).}
\centering
\includegraphics[width=0.95\textwidth]{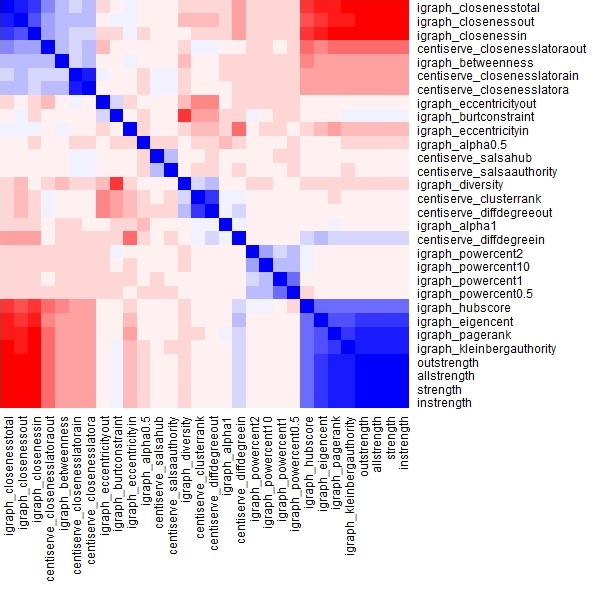}
\label{fig:TAcentralitycorr}
\end{figure}
Before comparing the centrality measure rankings, we first remove the industries from the rest of New Zealand (North and South of the Waikato Region), as these have much larger inputs/outputs than those broken down by TA within the Waikato Region and risk dominating the results. Furthermore, we are focused here on identifying important industries within the Waikato Region. Figure \ref{fig:TAcentralitycorr} shows a heatmap of the Kendall correlation coefficients \cite{Kendall} between all the centrality measures considered here, with the exception of the S-core decomposition. 

We find that overall the different eigenvector-based centrality measures are highly correlated, in particular the Kleinburg Authority, Kleinburg Hub, and PageRank, and that these are strongly driven by the strength (total inputs/outputs) of the industries. The different path-based measures are also highly correlated, for example, the different closeness and betweenness measures. These path-based measures highlight different industries to those identified by the eigenvector-based methods, as shown by the high level of anti-correlation (red) between these subsets of measures. More importantly, the path-based measures identify industries that would not be immediately revealed by eigenvector-based methods, or by traditional multiplier measures based on their size (strength) in the local economy. We elaborate on this in the next section.

If we aggregate the ten TAs up to a single Waikato Region, we can compare centrality measure correlations over a much smaller set of industry nodes (106 instead of 1060). This produces slightly stronger/weaker correlations, as shown in Figure \ref{fig:REGcentralitycorr}, but overall the pattern remains.

\begin{figure}[h]
\caption{Centrality measure correlations for industries in the Waikato Region as a single network region, colour scale from blue ($\tau$=-0.74) to red ($\tau$=1).}
\centering
\includegraphics[width=0.95\textwidth]{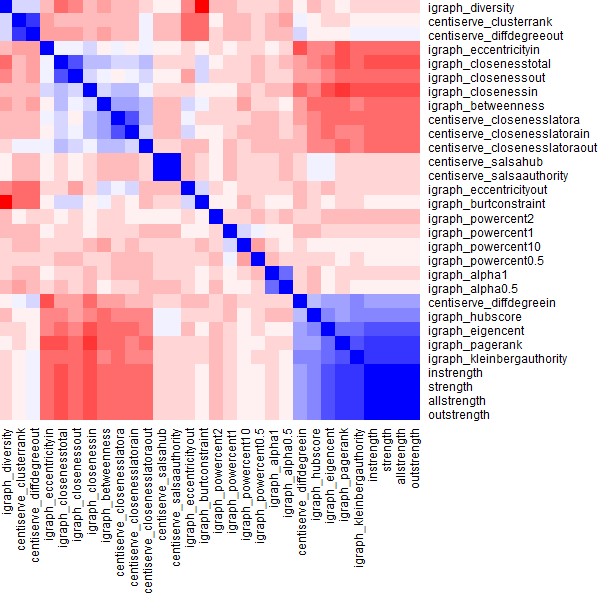}
\label{fig:REGcentralitycorr}
\end{figure}

% Put in a couple of figures that go on the pathway to being about to calculate the clusters of indices. Then jump to the indices. 

\subsection{Comparing multiplier and centrality measures}

Comparing rankings for the different multipliers calculated, we find that the correlation between the Industry (Type I) and Household (Type II) disruption multipliers is $\tau=0.21$. This matches the literature which shows that the inclusion of the household sector has a large impact on the results \cite{MillerBlair2009}. When we look at the Disruption multiplier, we find that this has a correlation of $\tau=0.38$ with the Industry multiplier and $\tau=-0.01$ with the Household multiplier. This shows that the Disruption model, which propagates a disruption through the IO network, is measuring a different disruption effect to the existing IO multiplier analysis. This has implications for regional disruption planning.

Comparing the three multipliers with the network centrality measures (Figure \ref{fig:TAMultiplierCorrelations}A), we find that the overall correlations are lower (-0.33 to 0.36) but that overall the eigenvector-based centralities and the overall industry strengths tend to match up with the traditional IO multipliers. This is not unexpected as they are both based on linear algebra matrix calculations that are mathematically similar, and the numerics agree with this.  More interestingly the path-based measures, such as the betweenness and the closeness, are much more strongly correlated with the Disruption multipliers. This makes intuitive sense as they are both concerned with flows and bottlenecks, that is with quantifying how disruptions to specific nodes flow on to impact the activity of dependent nodes.
Another point to note is that we find the having a high diversity score is connected to having both a high Industry multiplier and a high Disruption multiplier. This highlights  the potential importance of rarer industries within economic networks. % does it...??? I don't know. Throwaway comment...
Aggregating to a single Waikato Region, we find much the same results (Figure \ref{fig:TAMultiplierCorrelations}B), but with the disruption results matching more closely with a wider range of centralities. This is due to the disruption modelling becoming more homogenous in terms of industry distrubition and activity when looking at the aggregated Region. A feature of the Disruption model is that it was designed to consider lower levels of aggregation, with the aim to be able to provide detailed results at single industry level resolution.

\begin{figure}[t]
\caption{Comparing the network centrality measures with: the two IO multipliers, the mean of the IO multipliers, and the disruption multiplier. This shows correlations between eigenvector-based centrality measures and IO multipliers, whereas path-based centrality measures correlate well with Disruption multipliers. There is a negative correlation between the two.
For (A) industries in the ten TAs in the Waikato Region, colour scale from blue ($\tau$=-0.33) to red ($\tau$=0.36), and (B) industries in the Waikato Region as a whole, colour scale from blue ($\tau$=-0.76) to red ($\tau$=0.38).}
\centering
\includegraphics[width=0.5\textwidth]{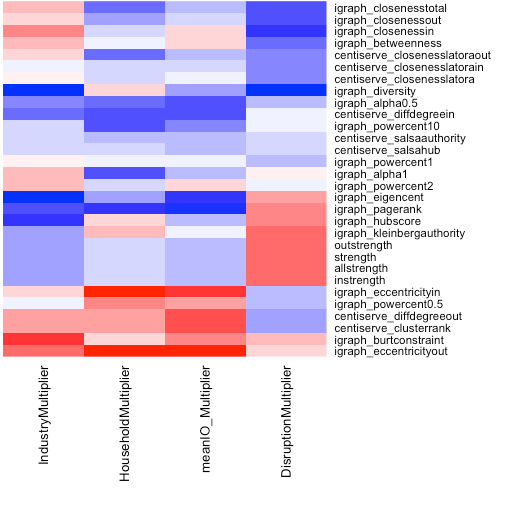}
\includegraphics[width=0.475\textwidth]{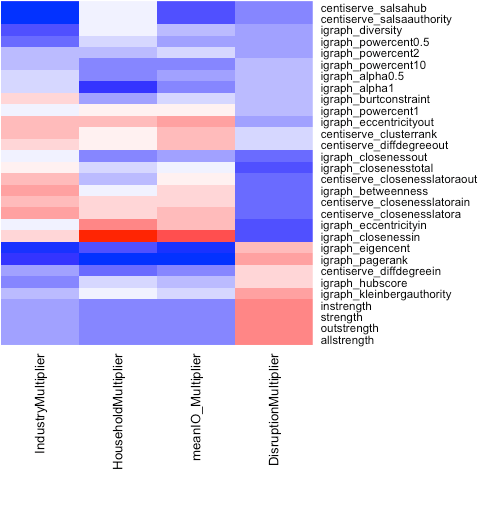}
\label{fig:TAMultiplierCorrelations}
\end{figure}

\subsection{Impact of spatial aggregation}
For all the analyses performed we considered the IO tables and economic networks with the Waikato Region broken down into 10 subregions (TAs) as well as with the whole Waikato Region considered together. This allowed us to look at the impact of the spatial aggregation on the industries identified as as important from the disruption analysis. 

We find that the eigenvector and strength based network measures identify the same key industries at both Region and TA levels of resolution. We find the same pattern for the Industry and Household multipliers, and the S-core decomposition. The value in analysing the system at the TA level disaggregation then becomes simply the ability to identify which TA the identified industry is most important to --- it does not change which industries are identified.
The exception is for industries that are disproportionately (or uniquely) represented in one or two TAs; for example, Coal Mining in the Waikato District, and Hospitals in Hamilton City. In these cases looking at TA level allows these to be ranked higher in importance than they would be if aggregating up to Regional level. Selected examples are given in Table \ref{tab:sameindustries}.

However, for any path-based measures and for the Disruption multipliers, the level of spatial aggregation has a large impact on which industries are identified as important. Examples are given in Table \ref{tab:differentindustries}. This can be explained as follows: aggregating the network up changes its structure --- for example, at TA level there are fewer individual businesses within each industry categorisation, so the self-loops are smaller. Furthermore, the proportion of inter-industry flows that are within the TA itself is quite low (14-32\%), with the majority of flows into (or out) of each industry coming from (or going to) other TAs within the Waikato Region and the rest of NZ. When considering the whole region, the proportion of inter-industry flows that stay within the region increases to around 60\%. This is still far below the equivalent proportions that are typically observed in the literature when looking at IO networks at a whole country level \cite{McNerney2009}. It is therefore worth noting that metrics that are applicable for national level analysis may not behave as expected when working with disaggregated regional data, such as that considered here.

{\renewcommand{\arraystretch}{1.2}%
\begin{table}[tbp]
\centering
\begin{tabular}{P{2.5cm}|@{\hspace{1em}} P{5cm} @{\hspace{1em}}| @{\hspace{1em}} P{5cm}}
\hline
Measure    & Regional level   & TA level    \\
\hline
Total industry value added & 
Dairy cattle farming\newline Owner-occupied property operation &  
Owner-occupied property operation (Hamilton City) \newline Hospitals (Hamilton City) 
\newline Dairy cattle farming (Matamata-Piako)
\\
\hline
Strength & 
Dairy product manufacturing \newline Dairy cattle farming \newline Electricity generation and on-selling &  
Dairy product manufacturing (3 TAs) \newline Electricity generation and on-selling (1 TA)
\newline Dairy cattle farming (1 TA) \newline Coal mining (Waikato District)
\\
\hline
PageRank & 
Dairy product manufacturing \newline Dairy cattle farming \newline Electricity generation and on-selling &  
Dairy product manufacturing (3 TAs) \newline Meat and meat product manufacturing (1 TA) \newline Dairy cattle farming (1 TA)
 \newline Coal mining (Waikato District)
\\
\hline
S-core decomposition (central subgraph) & 
Dairy product manufacturing 
\newline Dairy cattle farming 
\newline  Meat and meat product manufacturing 
\newline Building cleaning, pest control and other support services
\newline Non-residential property operation
\newline Hospitals
\newline Central government administration and justice
\newline Fruit, oil, cereal and other food product manufacturing
\newline Beverage and tobacco product manufacturing
\newline Rental and hiring services (except real estate); non-financial asset leasing
&  
\emph{same as Regional level}
\\
\hline
Industry IO multiplier & 
Electricity generation and on-selling \newline Primary metal and metal product manufacturing 
\newline Dairy product manufacturing \newline  Meat and meat product manufacturing &  
Electricity generation and on-selling (6 TAs) \newline Primary metal and metal product manufacturing  (7 TAs) \newline Dairy product manufacturing (7 TAs)
 \newline  Meat and meat product manufacturing (7 TAs)
\\
\hline
Household IO multiplier & 
Preschool education
 \newline Postal and courier pick up and delivery services
\newline Specialised food retailing  &  
Preschool education (10 TAs) \newline Postal and courier pick up and delivery services
  (10 TAs) \newline Specialised food retailing (10 TAs)
\\
\hline
\end{tabular}
\vspace*{5mm}
\caption{Selection of industries identified as important that are the same at Regional and TA level.}
\label{tab:sameindustries}
\end{table}

\begin{table}[tbp]
\centering
\begin{tabular}{P{2.5cm}|@{\hspace{1em}} P{5cm} @{\hspace{1em}}| @{\hspace{1em}} P{5cm}}
\hline
Measure    & Regional level   & TA level    \\
\hline
Betweenness & 
Petroleum and coal product manufacturing
 \newline Defence
 \newline Sewerage and drainage services
 \newline Air and space transport
  &  
Motor vehicle and motor vehicle parts wholesaling (Hauraki) \newline 
Sewerage and drainage services (Thames-Coromandel) \newline 
Waste collection, treatment and disposal services (2 TAs) \newline 
Health and general insurance (Taup{\=o}) \newline 
Petroleum and coal product manufacturing (Matamata-Piako) \newline
Warehousing and storage services (Waitomo)
\\
\hline
Closeness (total) & 
Air and space transport
 \newline Petroleum and coal product manufacturing
 \newline Beverage and tobacco product manufacturing
 \newline Warehousing and storage services
  &  
  Other transport (Otorohanga) \newline
Motor vehicle and motor vehicle parts wholesaling (Hauraki) \newline 
Warehousing and storage services (Waitomo)\newline 
Health and general insurance  (Taup{\=o}) \newline 
Waste collection, treatment and disposal services \newline 
Polymer product and rubber product manufacturing (2 TAs) \newline 
Machinery manufacturing \newline 
Electricity transmission and distribution
\\
\hline
Disruption multiplier & 
Defence
 \newline Life insurance
 \newline Petroleum and coal product manufacturing
 \newline Warehousing and storage services
  &  
Other transport (Otorohanga) \newline 
Motor vehicle and motor vehicle parts wholesaling (Hauraki) \newline 
Warehousing and storage services (Waitomo) \newline 
Waste collection, treatment and disposal services (Waitomo) \newline 
Polymer product and rubber product manufacturing (3 TAs) \newline 
Health and general insurance (Taup{\=o})
\\
\hline
\end{tabular}
\vspace*{5mm}
\caption{Selection of industries identified as important on measures that differ Regional and TA level.}
\label{tab:differentindustries}
\end{table}

\section{Discussion}
In this work we have considered the question of how to identify industries that have a large impact on an economic system when they are disrupted. In order to approach this, we have considered traditional IO multipliers, a new disruption model multiplier, and a range of network centrality measures. We have found that although traditional IO measures and eigenvector-based centrality measures are good at picking out the largest industries in terms of gross output or value-added, they do not match up with the industries that disruption modelling shows to have a large amplifying effect. We find that path-based measures, such as betweenness centrality, are far better at identifying industries that would have large flow-on impacts. These path-based methods explicitly consider the flow of money through the economy, and we find that the industries identified on these measures depend strongly on the level of spatial resolution.

In considering a natural hazard disruption, both the total size of the industry and the proportion it's impact gets amplified by will play a role in determining the resulting impacts. By taking a network science approach, we are able to get a fuller picture of the potential targets for mitigation investment (e.g.~stockpiling goods, having back-up generators in case of electricity outages).

In most disruption events, the impact will not be homogeneous through space. In most cases we would like to be able to consider the impact of a disruption on the well-being of communities, instead of just at national or even regional level. This is especially true for smaller localised events, which will not have a large impact at a national or regional level, but that could devastate a community. We have found that by considering smaller spatial units (in this case TA level) it is possible to get a better estimate of where the impacts will fall, as well as where susceptibilities are. Even for the measures that do not change much between Region and sub-regional (TA) level (Table \ref{tab:sameindustries}), looking at a higher granularity allows one to identify the unique (spatially specific) industries e.g.~Hospitals and Coal Mining, that would be missed at a Regional level. 

In future, it is foreseeable that increased data collection may make it possible to create networks at individual business level. Making sure that we understand how our analysis would scale from National to Regional all the way to individual level, will be an important focus of future research.

\paragraph{Acknowledgements}

Resilience to Nature's Challenges - National Science Challenge contestable funding 2017-2019. Ministry of Business, Innovation \& Employment, NZ. \url{http://resiliencechallenge.nz} \\
Te P{\=u}naha Matatini, Centre of Research Excellence, Tertiary Education Commission, NZ. \url{https://www.tepunahamatatini.ac.nz}

%
% ---- Bibliography ----
%

\end{document}